\documentclass[11pt]{article}
\usepackage{amsmath, latexsym,ifthen, amsfonts,amsbsy}
\usepackage{setspace}
\usepackage{subcaption}
\usepackage{natbib}
\usepackage{relsize}
\renewcommand{\harvardurl}[1]{\textbf{URL:} \url{#1}}
\usepackage{booktabs}
\usepackage{tkz-graph}
\usepackage{pgf}

\usepackage{setspace}
\usepackage{times}
\usepackage{lipsum}
\usepackage{tabularx}
   
\usepackage{amsmath,amsthm,amssymb,ifthen}
\usepackage{graphicx}
\usepackage[mathscr]{eucal}

\usepackage{enumerate}

\usepackage{etex}
\usepackage{booktabs}
\usepackage{amsfonts}
\usepackage[latin1]{inputenc}
\usepackage{soul,color}
\usepackage{chngcntr}
\usepackage{amsthm}
\usepackage{verbatim}
\usepackage{float}
\usepackage[hidelinks]{hyperref}
\usepackage{changepage}
\usepackage{amssymb}
\usepackage{adjustbox}

\usepackage{caption} 
\usepackage{xfrac}
\usepackage{graphicx}

\newtheoremstyle{note}
{8pt}
{8pt}
{}
{}
{\bfseries}
{:}
{.5em}
{}

\theoremstyle{note}
\newtheorem{theorem}{Theorem}
\newtheorem{lemma}[theorem]{Lemma}

\newtheorem{remark}{Remark}
\usepackage[T1]{fontenc}
\newtheorem{assumption}{Assumption}
\newtheorem{definition}{Definition}

\allowdisplaybreaks

\usepackage{multirow}
\usepackage[left=1in,right=1in,top=1in,bottom=1in]{geometry}
\usepackage{algorithm}
\date{}

\usepackage{soulpos}
\soulregister\cite7
\soulregister\citep7
\soulregister\ref7
\soulregister\emph7
\soulregister\eqref7
\soulregister\pageref7

\makeatletter
\newcommand{\vast}{\bBigg@{4}}
\newcommand{\Vast}{\bBigg@{5}}
\makeatother

\makeatletter
\newcommand*{\centernot}{%
  \mathpalette\@centernot
}
\def\@centernot#1#2{%
  \mathrel{%
    \rlap{%
      \settowidth\dimen@{$\m@th#1{#2}$}%
      \kern.5\dimen@
      \settowidth\dimen@{$\m@th#1=$}%
      \kern-.5\dimen@
      $\m@th#1\not$%
    }%
    {#2}%
  }%
}
\makeatother

\newcommand{\independent}{\perp\mkern-9.5mu\perp}
\newcommand{\notindependent}{\centernot{\independent}}

\def\bSig\mathbf{\Sigma}

\usepackage[figuresright]{rotating}

\usepackage{authblk}
\author[1]{BaoLuo Sun}
\author[2]{Wang Miao}
\affil[1]{ Department of Statistics and Applied Probability, National University of
Singapore}
\affil[2]{Guanghua School of Management, Peking University}

	{\raise0.5ex\hbox{$#1$}\! \left/ \! \lower0.5ex\hbox{$#2$}\right.}
	
\begin{document}

\title{On Semiparametric Instrumental Variable Estimation of Average Treatment Effects through Data Fusion}

	\clearpage \maketitle

\begin{abstract}	
	{	
Suppose one is interested in estimating causal effects  in the presence of potentially unmeasured confounding with the aid of a valid instrumental variable. This paper investigates the problem of making inferences about the average treatment effect when data are fused from two separate sources, one of which contains information on the treatment and the other contains information on the outcome, while values for the instrument and a vector of baseline covariates are recorded in both. We provide a general set of sufficient conditions under which the average treatment effect is nonparametrically identified from the observed data law induced by data fusion, even when the data are from two heterogeneous populations, and derive the efficiency bound for estimating this causal parameter. For inference, we develop both parametric and semiparametric methods, including a multiply robust and locally efficient estimator that is consistent even under partial misspecification of the observed data model. We illustrate the methods through simulations and an application on public housing projects.
	}
\end{abstract}	

\section{Introduction}
\label{sec:intro}

The instrumental variable method  is  widely used in the health and social sciences for identification and estimation of causal effects in the presence of potentially unmeasured confounding \citep{bowden1990instrumental, robins1994correcting, angrist1996identification,greenland2000introduction,wooldridge2010econometric,hernan2006instruments,didelez2010assumptions}. A valid instrumental variable $Z$ is a pre-exposure variable that is (a) associated with treatment $D$, (b) independent of any unmeasured confounder $U$ of the exposure-outcome association, and (c) has no direct causal effect on the outcome $Y$, conditional on a set of measured baseline covariates $X$. The instrumental variable approach has a longstanding tradition in econometrics going back to the original works of \cite{wright1928tariff} and \cite{goldberger1972structural} in the context of linear structural modeling; see \cite{wooldridge2010econometric}, \cite{clarke2012instrumental}, \cite{baiocchi2014instrumental} and \cite{swanson2018partial} for more recent reviews. Under correct specification of the linear structural equation models and assuming absence of baseline covariates, the conventional instrumental variable estimand of the average treatment effect is the population moment ratio $\text{cov}(Z,Y)/\text{cov}(Z,D)$.

However, in many empirical scenarios only information on $(Y,Z,X)$ is available from the primary population of interest. \cite{angrist1992effect} and \cite{10.2307/2297863} showed that the two sets of moments can be estimated from two separate sources by leveraging information on $(D,Z,X)$ from an auxiliary population, a method known as two-sample instrumental variable estimation. Furthermore, \citet{klevmarken1982missing} and \citet{angrist1995split} introduced two-sample two stage least squares estimation with first-stage  regression for the treatment model based on the auxiliary sample; see \citet{ridder2007econometrics} and \citet{angrist2008mostly} for reviews. This methodology has since been widely applied in econometrics and social sciences \citep{inoue2010two}, and more recently in two-sample Mendelian randomization studies to estimate causal relationships using genetic factors as instruments \citep{pierce2013efficient, gamazon2015gene, lawlor2016commentary, zhao2018statistical, zhao2019}. As noted by \citet{zhao2019}, the aforementioned methods typically assume that the auxiliary data is also sampled from the primary population. In addition, linear structural models impose strong homogeneity assumptions on the treatment effect. A robust analytic framework for instrumental identification and estimation of causal effects under data fusion therefore remains of keen interest in observational studies. \citet{graham2016efficient} identified the two-sample instrumental variable problem as one specific example of a general class of data combination models, and extended the semiparametric efficiency theory of \citet{hahn1998role} and \citet{chen2008semiparametric} to this class of models. Recent work has also made significant strides towards relaxing the assumptions for identification of causal effects under data fusion \citep{pacini2016robust, choi2018weak,zhao2018statistical,buchinsky2018estimation, shu2018improved,  zhao2019, pacini2019two}. 

 When full data on $L=(Y,D,Z,X)$ are available from the primary population of interest, \cite{robins1994correcting}, \cite{angrist1995identification}, \cite{angrist1996identification} and \cite{heckman1997instrumental} formalized the instrumental variable approach under the potential outcome framework \citep{neyman1923applications, rubin1974estimating}, which allows one to nonparametrically define the causal estimands of interest. In this paper, we propose novel assumptions under which the average treatment effect of $D$ on $Y$ in the primary population of interest can be uniquely and nonparametrically identified from the observed data law induced by data fusion. To estimate this identifying statistical functional, we develop a suite of parametric and semiparametric estimators including a multiply robust and locally efficient one that remains consistent even if the observed data model is partially misspecified. We compare the proposed estimators both in theory and via simulations, and investigate issues of efficiency and robustness of existing estimators.

\section{Model }

Suppose we are interested in estimating the average treatment effect of a binary treatment $D$ on outcome $Y$ in a primary population of interest, which is confounded by measured covariates $X$ as well as unmeasured ones $U$, with the aid of a binary instrumental variable $Z$. However, we only observe $\{(Y_i,Z_i,X_i)^T$, $i=1,...,n_p\}$ from this population. As a remedy, suppose an additional sample $\{(D_i,Z_i,X_i)^T$, $i=1,...,n_a\}$ is available from an auxiliary population, possibly different from the primary population. Similar to \cite{graham2016efficient,shu2018improved}, we assume the following about the  data source mechanism:
\begin{assumption}[Binomial sampling]
\label{datasource}
The combined set of $n=n_p+n_a$ units are independently drawn from either the primary population with a fixed probability $Q_0\in (0,1)$ or the auxiliary population with probability $1-Q_0$. 
\end{assumption}
Let $R_i$ be an indicator variable, equal to $1$ if the {\it i}th unit is drawn from the primary population, and $0$ otherwise. By assumption \ref{datasource}, the combined set of observed data $\{O_i=(R_i,R_iY_i,(1-R_i)D_i, Z_i,X_i)^T$, $i=1,...,n\}$ can be treated as a random sample from a synthetic merged population. Let $F(O)$ denote the distribution of $O$, with density with respect to some dominating measure given by
\begin{eqnarray}
f(O)=q^{\dag R}(1-q^{\dag})^{1-R}f(V|R=1)^Rf(V|R=0)^{1-R}f(Y|V,R=1)^Rf(D|V,R=0)^{1-R}, \label{eq:obslaw}
\end{eqnarray}
where $V=(Z,X)$ and $q^{\dag}=\text{pr}(R=1)$. Let $E(\cdot)$ denote expectation taken with respect to this mixture distribution, and let $\pi(z,x)=E(R|Z=z,X=x)$. By Bayes' rule,
$$f(z,x|R=1)=f(z,x|R=0)\left\{\frac{1-q^{\dag}}{q^{\dag}}\frac{\pi(z,x)}{1-\pi(z,x)}  \right\}.$$
Let $Y_d$ for $d\in \{0,1\}$ denote the potential outcome that would be observed if $D$ were set to $d$, which is related to the observed data via the consistency assumption $Y_d=Y$ if $D=d$.  To achieve identification of $\Delta\equiv E(Y_1-Y_0|R=1)$ based on the observed data law $F(O)$ induced by data fusion, we make the following assumptions about the primary and auxiliary populations.

\subsection{Primary population}

Suppose $Z$ is a valid binary instrument that satisfies the following assumptions \citep{didelez2007mendelian, pearl2009causality, clarke2012instrumental}:

\begin{assumption}[Instrument Relevance]
\label{ivrel}
$Z \notindependent D | X,R=1$.
\end{assumption}

\begin{assumption}[Instrument Independence]
\label{ivind}
$Z \independent U | X,R=1$.
\end{assumption}

\begin{assumption}[Exclusion Restriction]
\label{er}
$Y \independent Z | D,X,U,R=1$.
\end{assumption}

 Here $A\independent B|C$ indicates conditional independence of $A$ and $B$ given $C$ \citep{dawid1979conditional}. Instrument relevance ensures that $Z$ is a correlate of the
exposure even after conditioning on $X$, while instrument independence states that $Z$ is independent of all unmeasured confounders of the exposure-outcome association. Exclusion restriction formalizes the assumption of no direct
effect of $Z$ on $Y$ not mediated by $D$. Furthermore, the assumption of no unmeasured confounding given $(X,U)$ can be stated as 
\begin{assumption}[Latent Ignorability]
\label{li}
$Y_d \independent D|X,U,R=1$, $\text{for }d\in \{0,1\}$ \citep{robins1994correcting}.
\end{assumption}

Assumptions 2--5 may be known to hold at the design stage when the investigator controls treatment allocation conditional on baseline covariates in double blind randomized trials \citep{ten2008intent}. In observational studies, the potential instrumental variable may be viewed as being randomized through some natural or quasi-experiment within levels of the observed covariates \citep{hernan2006instruments}, although these assumptions are typically untestable without further conditions. The exclusion restriction assumption \ref{er} implies the following semiparametric structural models:
\begin{equation}
\begin{aligned}
 E(D\mid Z,X,U,R=1)&=g_0(X,U)+g_1(X,U)Z\\
 E(Y\mid D,Z,X,U,R=1)&=h_0(X,U)+h_1(X,U)D,\label{eq:outstruc}
\end{aligned}
\end{equation}
where  for $k\in\{0,1\}$, $g_k(\cdot)$ and $h_k(\cdot)$ are  arbitrary square-integrable functions of $(X,U)$ that are only restricted by natural features
of the model, e.g. such that the exposure mean is bounded between zero and one. Note that for binary $(Z,D)$, model (\ref{eq:outstruc}) is saturated as there are no restrictions on the corresponding data laws $f(D|Z,X,U,R=1)$ and $f(Y|D,Z,X,U,R=1)$ except for the implications of assumption \ref{er}. Under assumptions \ref{er} and \ref{li}, $h_{1}\left( x,u\right) =E(Y_1-Y_0|X=x,U=u,R=1)$ encodes the conditional average treatment effect within levels of $(X,U)$, hence $\Delta=E\left\{h_{1}\left( X,U\right) |R=1\right\}.$ The linear structural equation model \citep{wright1928tariff, goldberger1972structural}
\begin{equation}
\begin{aligned}
E(D\mid Z,X,U,R=1) &= \theta_0 +  \theta_1 X + \theta_2 U+ \theta_3 Z\\ 
E(Y\mid D,Z,X,U,R=1)&= \beta_0  +  \beta_1 X  +\beta_2 U+ \Delta D,\label{linear.eq}
\end{aligned}
\end{equation}
is a special case of (\ref{eq:outstruc}), where the function $h_1(X,U)$ is reduced to the scalar parameter of interest $\Delta$ encoding the homogeneous average treatment effect within levels of $(X,U)$. 

Even when full data on $L=(Y,D,Z,X)$ are available from the primary population, it is well known that while a valid instrumental variable satisfying assumptions \ref{ivrel}--\ref{li} suffices to obtain a valid statistical test of the sharp null hypothesis of
no individual causal effect, the population average treatment effect $\Delta$ is itself not uniquely identified from the law $F(L|R=1)$ \citep{balke1997bounds}. With a further monotonicity assumption about the effect of $Z$ on $D$,  \citet{angrist1996identification} showed that the local average treatment effect (LATE) among compliers can be nonparametrically identified. This framework has been further generalized in recent years by \cite{abadie2002instrumental}, \cite{abadie2003semiparametric}, \cite{carneiro2003understanding}, \cite{tan2010marginal}, \cite{ogburn2015doubly} and \cite{kennedy2019robust}.   \citet{zhao2019} discussed identification of LATE in two-sample instrumental variable analyses. However, because the population of compliers is itself nonidentifiable in general, $\Delta$ is arguably still a causal parameter of interest in many observational studies \citep{ 10.2307/2291630, imbens2010better}. \cite{wang2018bounded} proved identifiability of $\Delta$ from the law $F(L|R=1)$ under the additional assumption
\begin{eqnarray}
\label{wang}
 g_1(X,U)=g_1(X) \quad \text{or}\quad h_1(X,U)=h_1(X) \quad\text{ with probability 1,}
\end{eqnarray}
 i.e. at least one of these effects is not allowed to vary with $U$. We show that $\Delta$ can be identified from $F(O)$ provided $X$ must be sufficiently rich so that the effect of exposure on the outcome is uncorrelated with the effect of the instrument on the exposure conditional on $X$ \citep{cui2019semiparametric}, which can be achieved even if $X$ does not include all confounders of the effect of $D$ on $Y$.  
\begin{assumption}[Orthogonality]
\label{cov}
$\text{cov}\{g_1(X,U),h_1(X,U)|X,R=1\}=0 \quad{\text{with probability 1 }}.$
\end{assumption}
Assumption \ref{cov} may hold under certain data generating mechanisms even if (\ref{wang}) does not, and is guaranteed to hold under the sharp causal null effect. In addition, we require every unit within levels of the observed covariates to have some chance of receiving each level $z\in\{0,1\}$ of the instrument.
\edef\oldassumption{\the\numexpr\value{assumption}+1}
\renewcommand{\theassumption}{\oldassumption}
\begin{assumption}[Positivity]
\label{overlap} $0<\text{pr}(Z=1|X,R=1)<1$ with probability 1.
\end{assumption}

\subsection{Auxiliary population}

We make the following assumptions about the auxiliary population:
\edef\oldassumption{\the\numexpr\value{assumption}+1}
\renewcommand{\theassumption}{\oldassumption}
\begin{assumption}[Support overlap]
\label{overlap} $0<\pi(Z,X)<1$ with probability 1.
\end{assumption}
\edef\oldassumption{\the\numexpr\value{assumption}+1}
\renewcommand{\theassumption}{\oldassumption}
\begin{assumption}[Propensity score equality]
\label{equal}
$$\text{pr}(D=1|Z,X,R=0)=\text{pr}(D=1|Z,X,R=1)$$ with probability 1.
\end{assumption}
\edef\oldassumption{\the\numexpr\value{assumption}+1}
\renewcommand{\theassumption}{\oldassumption}
Assumption \ref{overlap} ensures that the support of the common variables $(Z,X)$ in the primary population is contained within that in the auxiliary population, and together with assumption \ref{equal} allows us to identify $\tau(z,x)$ based on $F(O)$. Assumption \ref{equal} only requires predictive invariance for the treatment between the two heterogeneous populations, and we do not require the stronger condition of ``structural invariance''  (e.g.  assumptions \ref{ivind}--\ref{cov} also hold in the auxiliary population), which is related to the notions of ``invariant prediction" \citep{peters2016causal}, ``autonomy" \citep{haavelmo1944probability} and ``stability" \citep{pearl2009causality} as discussed in \citet{zhao2019}.

\subsection{Nonparametric identification}

We  show that under assumptions 1--9, $\Delta$ is a functional on the nonparametric observed data statistical model $\mathcal{M}_{\text{np}}=\{F(O): F(O) \text{ unrestricted}\}$ of all regular laws $F(O)$ that satisfy the positivity and support overlap assumptions. In the following, let $\tau(z,x)=\text{pr}(D=1|Z=z,X=x,R=1)$ and $\lambda(z|x)=\text{pr}(Z=z|X=x,R=1)$ denote the treatment propensity score and probability density or mass function of $Z$ given $X$ respectively in the primary population.

\begin{theorem}
\label{thm:id}
Under assumptions 1--9, 
\begin{eqnarray}
\label{ident} \Delta = E \left\{\frac{R}{q^{\dag}}\frac{(-1)^{1-Z}}{\lambda(Z|X)}\frac{Y}{[\tau(1,X)-\tau(0,X)]}\right\}.\label{eq:ident}
\end{eqnarray}
\end{theorem}

\begin{remark}
 When $Y$ is continuous and $D$ and $Z$ are discrete of finite domain, the canonical instrumental variable assumptions 3 and 4 impose no constraints on the law $F(L|R=1)$ \citep{bonet2013instrumentality}. In addition, assumption 9 is akin to coarsening at random, which leaves the observed data law $F(O)$ unrestricted \citep{robins1997non, van2003unified}. When $Y$ is also discrete, assumptions 3 and 4 impose inequality constraints which do not restrict the parameter space of $F(L|R=1)$ locally if the true observed data law lies in the interior of the space defined by these constraints \citep{wang2017falsification, wang2018bounded}.
\end{remark}

\begin{remark}
While nuisance parameters such as $\{\lambda(\cdot), \tau(\cdot)\}$ can in principle be estimated nonparametrically using methods such as sieve estimation \citep{hahn1998role,hirano2003efficient,chen2008semiparametric}, in this paper we focus on parametric working models due to the curse of dimensionality when $X$ is of moderate or high dimension \citep{robins1997toward}. Since one cannot be confident that any of these models is correctly specified, we also propose an estimator of $\Delta$ that is robust to misspecifications of these models.
\end{remark}

\section{Estimation}

\subsection{Maximum likelihood estimation}

  Let $\hat{E}(\cdot)$ denote the empirical mean operator $\hat{E}\{ h(O)\} = n^{-1}\sum_{i=1}^n h(O_{i})$, and let $(\hat{\alpha},\hat{\psi},\hat{\xi}, \hat{\theta})$ denote the maximum likelihood estimators of $(\alpha,\psi,\xi,\theta)$ that index the parametric models $\pi(z,x;\alpha)$, $\lambda(z|x; \psi)$, $\tau(z,x;\xi)$ and additionally $f(y|z,x, R=1; \theta)=f(Y=y|Z=z,X=x, R=1; \theta)$ for the outcome conditional density specified by the analyst. We note that under assumption \ref{equal}, $\tau(z,x)=\text{pr}(D=1|Z=z,X=x,R=0)$ so that inferences on $\xi$ can be based on the auxiliary sample.  By taking iterated expectation of (\ref{eq:ident}) with respect to $(Z,X)$, the plug-in estimator of $\Delta$ is
\begin{eqnarray}
\hat{\Delta}_{\text{mle}}=\hat{E} \left\{\frac{1}{\hat{q}}\frac{(-1)^{1-Z}}{\lambda(Z|X;\hat{\psi})}\frac{\pi(Z,X;\hat{\alpha})E(Y|Z,X,R=1; \hat{\theta})}{\tau(1,X; \hat{\xi})-\tau(0,X; \hat{\xi})}\right\},
\end{eqnarray}%
where the distribution of $(Z,X)$ is estimated by its empirical distribution and $\hat{q}=\hat{E} (R)$. It is clear that consistency of $\hat{\Delta}_{\text{mle}}$ relies on correct specifications of the models $\pi(z,x;\alpha)$, $\lambda(z|x; \psi)$, $\tau(z,x;\xi)$ and $f(y|z,x,R=1;\theta)$. In the following we propose several semiparametric estimators of $\Delta$ that do not require these models to be fully specified. We proceed by first noting the following decomposition of the outcome conditional mean model.
\begin{lemma}
\label{lem:outcome}
Under assumptions 2--6,
\begin{eqnarray}
E(Y|Z=z,X=x,R=1)=\mathcal{H}(x)\tau(z,x)+\omega(x), \label{eq:outmean}
\end{eqnarray}
where $\omega(x)\equiv\text{cov}[g_1\left( X,U\right) , h_1\left( X,U\right) |X=x,R=1]+E[h_0(X,U)|X=x,R=1]$ and $\mathcal{H}(x)\equiv E[h_1(U,X)|X=x,R=1]$ is the treatment effect curve conditional on observed covariates. Therefore, $\Delta= E\{\mathcal{H}(X)|R=1\}$. 
\end{lemma}

\subsection{Semiparametric estimation}

Consider the following submodels of $\mathcal{M}_{\text{np}}$ in which smooth parametric models (indexed by finite-dimensional parameters) for certain components of the observed data law $F(O)$ are correctly specified:
\begin{definition}
		\item[$\mathcal{M}_1$]: The models $\lambda(z|x; \psi)$ and $\tau(z,x;\xi)$ are correctly specified such that $\lambda(z|x; \psi^{\dag})=\lambda(z|x)$ and $\tau(z,x;\xi^{\dag})=\tau(z,x)$ for some unknown values $(\psi^{\dag},\xi^{\dag})$; 
		\item[$\mathcal{M}_2$]: The models $\mathcal{H}(x;\gamma)$, $\omega(x;\eta)$ and  $\tau(z,x;\xi)$ are correctly specified such that $\mathcal{H}(x;\gamma^{\dag})=\mathcal{H}(x)$, $\omega(x;\eta^{\dag})=\omega(x)$ and $\tau(z,x;\xi^{\dag})=\tau(z,x)$ for some unknown values $(\gamma^{\dag},\eta^{\dag},\xi^{\dag})$;  
		\item[$\mathcal{M}_3$]: The models $\mathcal{H}(x;\gamma)$, $\omega(x;\eta)$ and  $\pi(z,x;\alpha)$ are correctly specified such that $\mathcal{H}(x;\gamma^{\dag})=\mathcal{H}(x)$, $\omega(x;\eta^{\dag})=\omega(x)$ and $\pi(z,x;\alpha^{\dag})=\pi(z,x)$ for some unknown values $(\gamma^{\dag},\eta^{\dag},\alpha^{\dag})$.
\end{definition}
We propose semiparametric estimators for $\Delta$ which are consistent and asymptotically normal in each of the above submodels.
The asymptotic variance formula of each estimator described in this section follows from standard M-estimation theory with estimated nuisance parameters \citep{newey1994large, van2000asymptotic}. Alternatively, bootstrapping methods may be used for variance estimation in practice. 

Our first estimator $\hat{\Delta}_1$ of $\Delta$ is motivated by identification formula (\ref{ident}) which does not require specification of an outcome model for $f(y|z,x,R=1)$, and solves
\begin{eqnarray}
\label{est1} 0 &=& \hat{E} \left\{ \mu_1(O;\Delta,  \hat{\psi}, \hat{\xi},\hat{q}) \right\}\equiv \hat{E}\left\{\frac{R}{\hat{q}}\frac{(-1)^{1-Z}}{\lambda(Z|X; \hat{\psi})}\frac{Y}{[\tau(1,X; \hat{\xi})-\tau(0,X; \hat{\xi})]}-{\Delta}\right\}.
\end{eqnarray}

\begin{remark}
The models for $\{\lambda(\cdot), \tau(\cdot)\}$ can be specified and estimated without access to the outcome data.  Estimation of $\Delta$ using $\hat{\Delta}_1$ could therefore be considered as part of a more objective analysis design in the sense that it mitigates potential for  ``data-dredging'' exercises when the outcome model is fully specified \citep{rubin2007design}.
\end{remark}

We propose two additional estimators of $\Delta$ which do not require a model for $\lambda(\cdot)$ but instead posit models $\mathcal{H}(X;\gamma)$ and $\omega(X;\eta)$ for components of the outcome conditional mean (\ref{eq:outmean}). Consider the semiparametric estimators $\hat{\Delta}_2$ and $\hat{\Delta}_3$ which solve 
\begin{eqnarray}
0=\hat{E} \left\{ \mu_2(O;\Delta,\hat{\gamma}_2, \hat{q}) \right\}\equiv \hat{E} \left\{\frac{R}{\hat{q}}[\mathcal{H}(X;\hat{\gamma}_2)-\Delta]  \right\}
\end{eqnarray}
and
\begin{eqnarray}
0=\hat{E}\left\{ \mu_3(O;\Delta,\hat{\gamma}_3, \hat{q}) \right\}\equiv \hat{E} \left\{\frac{R}{\hat{q}}[\mathcal{H}(X;\hat{\gamma}_3)-\Delta]  \right\},
\end{eqnarray}
respectively, where the estimators $\hat{\gamma}_2$ and $\hat{\gamma}_3$  are constructed in a way such that they are consistent in the submodels $\mathcal{M}_2$ and $\mathcal{M}_3$ respectively, as follows. Let $v(X)$ and $w(X)$ be analyst-specified vector functions of the same dimensions as $\gamma$ and $\eta$ respectively, for example $\{v(X),w(X)\}=\{\partial \mathcal{H}(X;\gamma)/ \partial \gamma,\partial \omega(X;\eta)/ \partial \eta\}$, and let $\mathcal{G}_{v,w}(X,Z)=\{v^T(X)Z,w^T(X)\}^T$ where $A^T$ denotes the transpose of $A$. Then let $(\hat{\gamma}_2,\hat{\eta}_2)$   be the joint solution to the estimating equation
\begin{eqnarray*}
0&=& \hat{E} \bigr\{\mathcal{G}_{v,w}(X,Z) \bigr\{R[Y-\mathcal{H}(X;\gamma)\tau(Z,X; \hat{\xi})-\omega(X;{\eta})]-(1-R)\mathcal{H}(X;\gamma)[D-\tau(Z,X; \hat{\xi})]\bigr\}\bigr\},
\end{eqnarray*}
while $(\hat{\gamma}_3,\hat{\eta}_3)$ jointly solve
\begin{eqnarray*}
0=  \hat{E}  \left\{ \mathcal{G}_{v,w}(X,Z) \biggr\{R[Y-\omega(X;{\eta})]-\frac{(1-R)\pi(Z,X;\hat{\alpha})}{1-\pi(Z,X;\hat{\alpha})}\mathcal{H}(X;\gamma)D  \biggr\} \right\}.
\end{eqnarray*}

\begin{lemma}
Under standard regularity conditions \citep{newey1994large}, the estimators $\hat{\Delta}_1$, $\hat{\Delta}_2$, and $\hat{\Delta}_3$ are consistent and asymptotically normal in submodels $\mathcal{M}_1$, $\mathcal{M}_2$ and $\mathcal{M}_3$, respectively.
\end{lemma}

\begin{remark}
To ensure that the proposed estimators of $\Delta$ lie between $-1$ and $1$ in the case of binary $Y$, following \citet{wang2018bounded} we can specify a model such as 
$$\mathcal{H}(X;{\gamma})=\tanh (\gamma^TX)=\frac{\exp(2\gamma^TX)-1}{\exp(2\gamma^TX)+1},$$
which guarantees that $\mathcal{H}(X;{\gamma})\in [-1,1]$. In addition, instead of the decomposition (\ref{eq:outmean}) for continuous $Y$,  \citet{wang2018bounded} provided a variation independent decomposition of the components in the likelihood $\{\text{pr}(Y=1|Z,X,R=1),\text{pr}(D=1|Z,X,R=1)\}$ for binary $Y$, and their estimation strategy for these components may be adopted similarly. 
\end{remark}

\subsection{Multiply robust estimation}

To motivate the multiply robust estimator, we consider efficient estimation of $\Delta$ in $\mathcal{M}_{\text{np}}$. Any regular and asymptotically linear estimator $\hat{\Delta}$ has an associated influence function $\mu(O;\Delta)$ such that $\hat{\Delta}-\Delta=\hat{E}\{ \mu(O;\Delta)\}+o_p(n^{-1/2})$ \citep{bickel1993efficient}. Therefore it suffices to identify $\mu(O;\Delta)$ with the lowest variance, which is the efficient influence function.

\setcounter{theorem}{1}

\begin{theorem}
\label{thm:eff}
The efficient influence function for $\Delta$ in $\mathcal{M}_{\text{np}}$ is 
\begin{eqnarray}
\mu_{\text{\normalfont eff}}(O;\Delta) &=&  \frac{\left( -1\right)^{1-Z}\left\{ 
\begin{array}{c}
 \frac{R}{q^{\dag}}[ Y-\mathcal{H}\left( X\right)
\tau(Z,X)-\omega(X)]\\ 
-\frac{1-R}{q^{\dag}}\frac{\pi(Z,X)}{1-\pi(Z,X)}\mathcal{H}\left( X\right)[D-\tau(Z,X)] 
\end{array}%
\right\}}{\lambda(Z|X)\left[ \tau(1,X)-\tau(0,X)\right] }+\frac{R}{q^{\dag}}\{\mathcal{H}\left( X\right)-\Delta\},
\end{eqnarray}
so that the semiparametric efficiency bound for estimating $\Delta$ in $\mathcal{M}_{\text{np}}$ is $E\{\mu^2_{\text{\normalfont eff}}(O;\Delta)\}$.
\end{theorem}
We use $\mu_{\text{\normalfont eff}}(\cdot)$ as an estimating function and plug in estimates of the nuisance parameters to estimate the causal effect $\Delta$. This method of constructing estimating equations from influence functions has been used widely, e.g. in \cite{bang2005doubly, tan2006regression, tchetgen2009doubly, sun2016semiparametric, sun2018inverse, wang2018bounded}. Consider $(\tilde{\gamma}, \tilde{\eta})$ which jointly solve
\begin{equation}
\begin{aligned}
 \mathbf{0} &= &  \hat{E} \Biggr \{ \mathcal{G}_{v,w}(X,Z) \biggr\{R[Y-\mathcal{H}(X;\gamma)\tau(Z,X; \hat{\xi})-\omega(X; {\eta})]\\
&\phantom{=}&-\frac{(1-R)\pi(Z,X;\hat{\alpha})}{1-\pi(Z,X;\hat{\alpha})}\mathcal{H}(X;\gamma)[D-\tau(Z,X; \hat{\xi})]  \biggr\} \biggr\}. \label{dr1}
\end{aligned}
\end{equation}

We note that the estimator $\tilde{\gamma}$ is doubly robust in the sense that it is consistent for $\gamma^{\dag}$ in the model $\mathcal{M}_{2}\cup \mathcal{M}_{3}$, which is necessary for the multiply robust result stated below.
	\setcounter{theorem}{3}
\begin{lemma}
\label{lem:est5}
Under standard regularity conditions \citep{newey1994large}, the estimator $\hat{\Delta}_{\text{\normalfont mul}}$ which solves
\begin{eqnarray}
0&=& \hat{E}\left\{\mu_{\text{\normalfont eff}}(O; \Delta, \tilde{\eta},\tilde{\gamma},\hat{\psi},  \hat{\xi},\hat{\alpha},\hat{q})\right\}
\end{eqnarray}%
\begin{sloppypar}
\noindent is consistent and asymptotically normal in the union model $ \mathcal{M}_{\text{{\normalfont union}}}=\cup_{j=1}^3  \mathcal{M}_{j}$ (multiply robust). Moreover, $\hat{\Delta}_{\text{\normalfont mul}}$ attains the semiparametric efficiency bound in $\mathcal{M}_{\text{np}}$ (and, following the general results of \cite{robi}, also in  $\mathcal{M}_{\text{{\normalfont union}}}$) at the intersection submodel $\{\cap_{j=1}^3 \mathcal{M}_j\}$ where all working models are correctly specified (locally efficient).
 \end{sloppypar}
\end{lemma}

\section{Comparison to some existing estimators }
Suppose that $E(U|Z=z,X=x,R=1)=E(U|X=x,R=1)$ is linear in $x$, then the linear structural models (\ref{linear.eq}) yield the observed data models
\begin{eqnarray*}
\tau_{\text{linear}}(Z,X;\xi) &=&  \xi^T (1,Z,X)^T;\\
\quad \omega_{\text{linear}}(X;\eta)&=&\eta^T (1,X)^T;\\
 E(Y\mid Z,X,R=1)&=& \Delta \tau_{\text{linear}}(Z,X;\xi)+\omega_{\text{linear}}(X;\eta).\label{linearobs.eq}
\end{eqnarray*}
We also have that $\mathcal{H}\left(X\right)$ is indexed by the scalar parameter of interest $\Delta$. Using the notation in section 3, it can be shown that the two-sample instrumental variable estimator \citep{inoue2010two} $(\hat{\Delta}_{\text{tsiv}}, \hat{\eta}_{\text{tsiv}})$ solves
\begin{eqnarray*}
0=  \hat{E}  \biggr\{ \mathcal{G}_{v,w}(X,Z) \biggr\{R[Y-\omega_{\text{linear}}(X;\eta)]-\frac{(1-R)\hat{q}}{1-\hat{q}}\Delta D  \biggr\} \biggr\}.
\end{eqnarray*}
Inferences based on the two-sample instrumental variable estimator can be viewed as special instances of inferences obtained under a particular specification of submodel $\mathcal{M}_3$ with the above parametric models for $\{\mathcal{H}(\cdot),\omega(\cdot)\}$ and additionally $\pi(z,x;\alpha)=q$ where $q \in \mathbb{R}$, e.g. the marginal distribution of $(Z,X)$ is the same in the primary and auxiliary populations.  Therefore $\hat{\Delta}_{\text{tsiv}}$ will fail to be consistent for $\Delta$ if any of the parametric models in $\mathcal{M}_3$ is incorrectly specified. Furthermore, we note that the  two-sample two stage least squares estimator $(\hat{\Delta}_{\text{ts2sls}}, \hat{\eta}_{\text{ts2sls}})$ solves 
\begin{eqnarray*}
0&=&  \hat{E}  \biggr\{\mathcal{G}_{v,w}(X,Z)\biggr\{R[Y-\Delta\tau_{\text{linear}}(Z,X; \hat{\xi})-\omega_{\text{linear}}(X;\eta)]\\
&-&\frac{(1-R)\hat{q}}{1-\hat{q}}\Delta[D-\tau_{\text{linear}}(Z,X; \hat{\xi})]  \biggr\} \biggr\}, \label{eq:dr}
\end{eqnarray*}
 which is a special case of the doubly robust estimating equation (\ref{dr1}). It follows that $\hat{\Delta}_{\text{ts2sls}}$ is  consistent for $\Delta$ in $\mathcal{M}_2\cup\mathcal{M}_3$; even when the true marginal distribution of $(Z,X)$ differs between the primary and auxiliary populations, $\hat{\Delta}_{\text{ts2sls}}$ is  consistent provided the linear propensity score model $\tau_{\text{linear}}(\cdot)$ is correctly specified. We can also show via semiparametric effciency theory that $\hat{\Delta}_{\text{ts2sls}}$ is asymptotically more efficient than its non-doubly robust counterpart $\hat{\Delta}_{\text{tsiv}}$ at the intersection submodel $\mathcal{M}_2\cap\mathcal{M}_3$ \citep{tan2007comment, tsiatis2007semiparametric}.  The above properties were noted by \citet{inoue2010two}.
 
\citet{shu2018improved} proposed a class of doubly robust estimators $(\hat{\Delta}_{\text{dr}}, \hat{\eta}_{\text{dr}})^T$  which solve
\begin{eqnarray*}
0&=&  \hat{E}  \biggr\{\mathcal{G}_{v,w}(X,Z)\biggr\{R[Y-\Delta \tau(Z,X; \hat{\xi})-\omega_{\text{linear}}(X;\eta)]\nonumber\\
&-&\frac{(1-R)\pi(Z,X;\hat{\alpha})}{1-\pi(Z,X;\hat{\alpha})}\Delta[D-\tau(Z,X; \hat{\xi})]  \biggr\} \biggr\}, \label{eq:dr}
\end{eqnarray*}
where users can freely specify models for $\{\tau(\cdot), \pi(\cdot)\}$. \citet{graham2016efficient} introduced in earlier work a doubly robust auxiliary-to-study tilting estimator under restricted nuisance model specifications in efficient estimation of data combination models. Inferences based on $\hat{\Delta}_{\text{dr}}$ can be viewed as special instances of inferences obtained under a particular specification of submodel $\mathcal{M}_2\cup\mathcal{M}_3$ with $\mathcal{H}\left( X\right)=\Delta$ and $\omega_{\text{linear}}(\cdot)$. In constrast to $\hat{\Delta}_{\text{mul}}$, $\hat{\Delta}_{\text{dr}}$ will generally fail to be consistent for $\Delta$ outside the union model $\mathcal{M}_2\cup\mathcal{M}_3$. We note that a generalized version of $\hat{\Delta}_{\text{dr}}$ that accommodates arbitrary parametric model specifications in $\mathcal{M}_2\cup\mathcal{M}_3$ is given by 
\begin{eqnarray}
\hat{\Delta}_{\text{dr2}}= \hat{E} \left\{R\mathcal{H}(X;\tilde{\gamma})/\hat{q}  \right\}, \label{gendr}
\end{eqnarray}
 where $\tilde{\gamma}$ solves (\ref{dr1}).

\section{Simulation study}

We investigate the finite-sample properties of the proposed semiparametric estimators under a variety of settings. For the primary population, baseline covariates ${X}=(X_1,X_2,X_3)^T$ are mutually independent and marginally distributed as $\text{U}(0,1)$; $(Y,A,Z,U)$ is distributed as follows:
\begin{eqnarray*}
U|{X} &\sim& \text{TN}\{\vartheta^{T}{X},1,(\vartheta^{T}{X}-1,\vartheta^{ T}{X}+1)\}; \\
Z|{X} &\sim& \text{Bernoulli }\{p=\{1+\exp{[-{\psi}^{ T}(1,{X}^T)^T]}\}^{-1}\};\\
D|Z,{X},U&\sim&  \text{Bernoulli }\{p=\{1+\exp{[-{\xi}^{ T}(1,Z,{X}^T)^T]}\}^{-1}+0.2[U -\vartheta^{ T}X]\} ;\\
Y|D,{X},U&\sim& \text{N}\{\gamma^{T}(1,{X}^T)^TD+1.25\times\vec{1}^{\,T}{X}+6U, 1\},
\end{eqnarray*}%
where $\text{TN}\{\mu, \sigma^2,(l,u)\}$ denotes a truncated normal distribution with support $[l,u]$, $\vartheta=(0.5, -0.5,0)^T$, ${\psi}=(-1,0.5,0.5,0.5)^T$, ${\xi}=(-1.3,1.2,0.5,-0.25-0.25)^T$, $\gamma=(2,0.5,0.5,0.5)^T$ and $\vec{1}=(1,1,1)^T$. For the auxiliary population,  ${X}=(X_1,X_2,X_3)^T$ are mutually independent and marginally distributed as $\text{TN}\{0.5,1,(0,1)\}$, $Z|{X} \sim \text{Bernoulli }\{p=\{1+\exp{[-{\psi}^{ T}(1,{X}^T)^T]}\}^{-1}\}$ and $D|Z,{X}\sim  \text{Bernoulli }\{p=\{1+\exp{[-{\xi}^{ T}(1,Z,{X}^T)^T]}\}^{-1}\}$; the remaining parts of the data law are left unrestricted. For each simulation replicate of total sample size $n$, we generate $n_p\sim \text{binomial}(n,p=0.7)$, followed by an i.i.d. sample of size $n_p$ from the primary population with only realizations of $(Y,Z,X)$ recorded, and another i.i.d. sample of size $n_a=n-n_p$ from the auxiliary population with only realizations of $(D,Z,X)$ recorded. The two samples are then merged, and an indicator variable $R$ is introduced, equal to 1 or 0 if the unit is drawn from the primary or auxiliary population respectively. It can be verified that the above data generating mechanism satisfies assumptions 1--9, and that the corresponding true observed data models are $\lambda(1|x;{\psi})=\{1+\exp{[-{\psi}^T(1,x^T)^T]}\}^{-1}$, $\tau(z,x;\xi)=\{1+\exp{[-{\xi}^T(1,z,x^T)^T]}\}^{-1}$, $\mathcal{H}(x;\gamma)={\gamma}^T(1,x^T)^T$, $\omega(x;\eta)=\eta^T (1,x^T)^T$ and $\pi(z,x;\alpha)=\{1+\exp{[-{\alpha}^T(1,z,x^T,x^{2 T})^T]}\}^{-1}$ where $x^2=(x^2_1,x^2_2,x^2_3)^T$ (by Bayes' rule). We are interested in estimating the average treatment effect $\Delta=E\{\gamma^{ T}(1,X^T)^T|R=1\}=2.75$. The four semiparametric estimators $\hat{\Delta}_1$, $\hat{\Delta}_2$, $\hat{\Delta}_3$ and $\hat{\Delta}_{\text{\ mul}}$ are implemented using $v(x)=w(x)=(1,x^T)^T$ as index functions.

Similar to \cite{kang2007demystifying}, we evaluate the performance of the proposed estimators in situations where
some models may be mis-specified by considering the transformed variables $V^{\ast}=(Z^{\ast},X^{\ast}_1,X^{\ast}_2,X^{\ast}_3)^T$ where $Z^{\ast}\sim \text{Bernoulli}\{p=\Phi(-2+3Z)\}$, $X^{\ast}_1=\exp(-0.5X_1)+\epsilon_1$, $X^{\ast}_2=X_2/[1+\exp(Z)]+\epsilon_2$ and $X^{\ast}_3=(X_1X_3)^3+\epsilon_3$; $\Phi(\cdot)$ is the cumulative distribution function of the standard normal distribution and the error terms are generated as $(\epsilon_1,\epsilon_2,\epsilon_3)^T\sim N({0},{I}_3)$. Then a particular component model is mis-specified when the analyst uses $V^{\ast}$ instead of $V$ in the working model. Specifically, we report results from the following five scenarios:
\begin{description} 
	\item[$\mathcal{M}_0^{\prime}$:] All models are correct;
	\item[$\mathcal{M}_1^{\prime}$:]  Only models  $\lambda(z|x;{\psi})$ and $\tau(z,x;\xi)$  are correct;
	\item[$\mathcal{M}_2^{\prime}$:]  Only models $\tau(z,x;\xi)$, $\mathcal{H}(x;\gamma)$ and $\omega(x;\eta)$ are correct;
	\item[$\mathcal{M}_3^{\prime}$:] Only models $\pi(z,x;\alpha)$, $\mathcal{H}(x;\gamma)$ and $\omega(x;\eta)$ are correct;
\item[$\mathcal{M}_4^{\prime}$:] All models are incorrect.
\end{description}
All simulation results are based on 1000 Monte Carlo runs of $n=10000$ units each. Table \ref{tab:est} summarizes simulation results. In agreement with theory,  $\hat{\Delta}_1$ has small bias  in $\mathcal{M}_0^{\prime}$ and $\mathcal{M}_1^{\prime}$, $\hat{\Delta}_2$ has small bias in $\mathcal{M}_0^{\prime}$ and $\mathcal{M}_2^{\prime}$, $\hat{\Delta}_3$ has small bias in  $\mathcal{M}_0^{\prime}$ and $\mathcal{M}_3^{\prime}$, and $\hat{\Delta}_{\text{mul}}$ has small bias in $\mathcal{M}_l^{\prime}$, $l=0,1,2,3$. In $\mathcal{M}_0^{\prime}$ where all models are correct, $\hat{\Delta}_1$ and $\hat{\Delta}_2$  have smaller Monte Carlo standard errors compared to $\hat{\Delta}_3$  which involves weighting through the data source propensity score $\pi(z,x)$.

\begin{table}
	\begin{center}
		\small
		\caption{Monte Carlo results of the proposed semiparametric estimators under different scenarios}
		\bigskip
		\label{tab:est}
		\renewcommand{\arraystretch}{0.8}
		\begin{tabular}{ccccccccccccc}
			\toprule
	     \multirow{ 2}{*}{Model} &  \multicolumn{4}{c}	{Estimator}   \\[2pt]
	     	 \cmidrule(l){2-5} 
			& $\hat{\Delta}_1$ & $\hat{\Delta}_2$  &$\hat{\Delta}_3$ & $\phantom{=}\hat{\Delta}_{\text{mul}}$ \\
			\midrule
			\multicolumn{3}{l}{$\mid$Bias$\mid$ (SE) }	&&  \\[2pt]
 $\mathcal{M}_0^{\prime}$  & 0.01 (0.29) & 0.01 (0.29) & 0.08 (0.33) & 0.04 (0.31) \\ 
$\mathcal{M}_1^{\prime}$  & 0.01 (0.29)  & 0.65 (0.34) & 0.74 (0.37) & 0.05 (0.30) \\ 
$\mathcal{M}_2^{\prime}$ & 0.67 (0.36)  &  0.01 (0.32) & 0.11 (0.41) & 0.05 (0.33) \\
$\mathcal{M}_3^{\prime}$& 1.10 (0.46) &  1.20 (0.48) & 0.09 (0.34) & 0.06 (0.33)  \\ 
$\mathcal{M}_4^{\prime}$& 1.30 (0.47) & 2.20 (0.57) & 0.77 (0.44) & 0.72 (0.39)  \\ 
			\multicolumn{3}{l}{RMSE}	&& \\[2pt]
$\mathcal{M}_0^{\prime}$&  0.09  & 0.09 & 0.11 & 0.10\\
$\mathcal{M}_1^{\prime}$&  0.08  & 0.54 & 0.68 & 0.09  \\
$\mathcal{M}_2^{\prime}$& 0.58  & 0.10 & 0.18 & 0.11  \\ 
$\mathcal{M}_3^{\prime}$& 1.50 & 1.70 & 0.12 & 0.11 \\ 
$\mathcal{M}_4^{\prime}$& 1.80 & 5.00 & 0.78 & 0.67  \\ 
			\bottomrule 
		\end{tabular}
	\end{center}
\end{table}

\section{Application}

\citet{currie2000public} study the the effect of public housing participation on housing quality
and educational attainment, and showed that project participation is associated with poorer outcomes based on data from the Survey of Income and Program Participation (SIPP). However, many unobserved factors such as social ties are likely to affect both project participation and outcomes, and the authors suspect that failure to control for this source of endogeneity would bias the estimated causal effects of living
in projects downwards, since families in projects may be more likely to live in
substandard housing in any case, and their children may be more likely to
experience negative outcomes. Leveraging on the sex composition of children as an instrumental variable for project participation, \citet{currie2000public} use two-sample instrumental variable methods to combine information from the 1990 Census data and 1990-1995 waves of the March Current Population Survey (CPS), and find that project households are less likely to suffer from overcrowding or live in high-density complexes, and project children are less likely to have been held back. Their study is important as the results overturn the stereotype that project participation is harmful in terms of living conditions and children's educational attainment.

In this analysis, we apply the proposed methods to estimate the causal effect of project participation ($D$) on reported monthly rental payments ($Y$) in the SIPP population; reported rent may be viewed as a proxy for housing quality \citep{currie2000public}. The binary instrumental variable $Z$ takes on value $1$ if a family had a boy and a girl, and $0$ if both are boys or girls. Families with two children of opposite genders will be eligible for three-bedroom apartments as opposed to two-bedroom apartments, and therefore will be more likely to participate in the housing project, although there is little reason to expect that the children's sex composition will directly affect $Y$. In line with the \citet{currie2000public} study, the vector of baseline covariates $X$ include the household head's gender, age, race, education, marital status and the number of boys in the family. We specify main effects models for $\{\lambda(\cdot),\tau(\cdot),\pi(\cdot)\}$ with logistic links. In addition, following \citet{shu2018improved} we add an additional interaction term involving household head information to  the linear predictor function of the model for $\pi(\cdot)$ to improve covariate balance, and specify $\omega(x;\eta)=\eta^T (1,x^T)^T$, $\mathcal{H}(x;\gamma)={\Delta}$. The analysis results based on $n_1=116901$ renters' complete records for $(Y,Z,X)$ from the 1990 Census of SIPP $(R=1)$ and $n_0=10382$ renters' complete records for $(D,Z,X)$ from CPS $(R=0)$, for a total sample size of $n=127283$, are summarized in Table \ref{tab:2}. 

\begin{table}
\caption{{Estimates of the effect of public housing project participation on reported monthly rental (divided by 1000 US dollars).}}
\label{tab:2}\par
\vskip .2cm
\centering
\renewcommand{\arraystretch}{0.8}
\begin{tabular}{c c c c c c c c c c c c c c c c c}
\hline\noalign{\smallskip}

&point estimate & standard error& 95\% Wald CI \\
\hline\noalign{\smallskip}
\noalign{\smallskip}
\phantom{==}$\hat{\Delta}_{\text{ts2sls}}$ & 0.3717 & 0.1124  &  (0.1513, 0.5920) \\ 
$\hat{\Delta}_1$ & 0.7650&0.3442& (0.0903, 1.4397)\\
$\hat{\Delta}_2$ &0.3790&0.1162&(0.1513,  0.6068)\\
$\hat{\Delta}_3$ &0.4999&0.2533&(0.0034, 0.9964)\\
\phantom{=}$\hat{\Delta}_{\text{mul}}$ &0.9155&0.4126&(0.1069, 1.7242)\\
\hline
\end{tabular}
\end{table}

The  two-sample two-stage least squares estimate of $0.3717$ agrees with the point estimate presented in Table 4 of  \citet{currie2000public}, although the analytic standard error of $0.1124$ is larger than the value of $0.0589$ reported by the original study, as the former takes into account the variability associated with the first-stage estimation. While the point estimates of the proposed estimators are all larger than $0.3717$, the point estimate of $\hat{\Delta}_{\text{mul}}$ is closest to that of $\hat{\Delta}_1$, which suggests that the models for $\{\lambda(\cdot),\tau(\cdot)\}$ in this illustrative analysis may be specified nearly correctly; \citet{tchetgen2010semiparametric} describe a formal specification test to detect which of the  baseline models is correct under the union model $\mathcal{M}_{\text{{\normalfont union}}}$. The point estimate of $0.9155$ for $\hat{\Delta}_{\text{mul}}$ also suggests that the causal effect of housing project participation on improving household living conditions is probably larger than the value reported in \citet{currie2000public}, since  $\hat{\Delta}_{\text{ts2sls}}$ is generally no longer consistent outside the union model $\mathcal{M}_{2}\cup\mathcal{M}_3$.

\section{Discussion}
\label{sec:conc}

Suppose we observe data on $(D,Z,X)$ from the primary population of interest and fuse it with data on $(Y,Z,X)$ from an auxiliary source,  i.e. $R_i$ equals to either $0$ or $1$ if the {\it i}th unit is drawn from the primary or the auxiliary population respectively. In this case, it is clear that inference about the identifying functional 
\begin{eqnarray*}
 \Delta = E \left\{\frac{1-R}{1-q^{\dag}}\frac{(-1)^{1-Z}}{\lambda(Z|X)}\frac{Y}{[\tau(1,X)-\tau(0,X)]}\right\}
\end{eqnarray*}
 is not possible under submodel $\mathcal{M}_1$, since $Y$ is not observed from the primary population. Nonetheless,  inference for $\Delta$ is still possible under $\mathcal{M}_2\cup \mathcal{M}_3$ if we replace assumption \ref{equal} with predictive invariance for the outcome: 
\begin{assumption}
\label{outequal}
$E(Y|Z,X,R=0)=E(Y|Z,X,R=1)$ with probability 1.
\end{assumption}
Indeed, it can be shown that under assumptions 1--8 and \ref{outequal}, the estimator 
\begin{eqnarray}
\tilde{\Delta}_{\text{dr3}}= \hat{E} \left\{(1-R)\mathcal{H}(X;\tilde{\gamma})/(1-\hat{q})  \right\}, \label{gendr}
\end{eqnarray}
 where $\tilde{\gamma}$ solves (\ref{dr1}) is consistent and asymptotically normal  in the union model $\mathcal{M}_2\cup\mathcal{M}_3$. We note that because $\hat{\Delta}_{\text{tsiv}}$, $\hat{\Delta}_{\text{ts2sls}}$ and $\hat{\Delta}_{\text{dr}}$ typically specify $\mathcal{H}\left( x;\gamma\right)=\Delta$ which does not depend on values for the baseline covariates, one can be agnostic as to which of the two samples is drawn from the primary population as long as assumptions 1--10 all hold. 

\begin{sloppypar}
There are several improvements and extensions for future work. Multiple valid instrumental variables can be incorporated by adopting a standard generalized
method of moments approach \citep{hansen1982large}, and the proposed estimators can be improved in terms of efficiency \citep{tan2006distributional, tan2010nonparametric} and bias \citep{vermeulen2015bias}. In this paper, we focused on the canonical case of binary $Z$ and $D$; extension of the proposed methodology to the case of general $Z$ or $D$ is an interesting topic for future research. It will also be of interest to investigate the use of negative controls under data fusion to mitigate unmeasured confounding and identify causal effects, which  has gained increasing recognition and popularity in recent years \citep{miao2017invited, shi2018multiply}.
\end{sloppypar}

In settings where $X$ is high dimensional, various flexible and highly data-adaptive machine learning methods may be adopted to estimate the nuisance parameters $\delta=\{\omega(\cdot), \mathcal{H}(\cdot), \lambda(\cdot),\tau(\cdot),\pi(\cdot)\}$, including random forests, lasso or post-lasso, neural nets or ensembles of these methods. This is useful for example if one does not wish to impose parametric models for the conditional treatment effect curve $\mathcal{H}(\cdot)$, or for $\pi(\cdot)$ which encodes the differences in the marginal distributions of $(Z,X)$ between the primary and auxiliary populations. By exploiting a condition known as Neyman orthogonality \citep{neyman1,neyman2, belloni2017program, 10.1111/ectj.12097} which translates to reduced sensitivity under local variation in the nuisance parameter, \citet{10.1111/ectj.12097}  show that $n^{-1/2}$ consistent estimation of $\Delta$ is possible under rate conditions for estimation of ${\delta}$ even when the complexity of the nuisance model space is no longer limited by classical settings, e.g. Donsker classes. Assume that the estimator $\hat{\delta}$ based on data of sample size $n$ takes values in $\mathcal{T}_n\subset \mathcal{T}$ with high probability,  where $\mathcal{T}$ is the set of all $\delta$ consisting of square-integrable functions. The proof for Lemma \ref{lem:est5} can be extended to show that  the estimating function $\mu_{\text{\normalfont eff}}(\cdot)$ satisfies the Neyman orthogonality property with respect to the nuisance realization set $\mathcal{T}_n$. The impact of regularization bias and overfitting in estimation of $\delta$ is further mitigated via cross-fitting \citep{10.1111/ectj.12097}. The performance of the resulting cross-fitted debiased machine learning estimators of $\Delta$ is intimately tied to the performance of the nuisance parameter estimator $\hat{\delta}$.  \citet{cui2019biasaware} recently introduced a framework for selective machine learning estimation  based on minimization of a certain cross-validated quadratic pseudo-risk, which may be adopted here by leveraging the multiple robustness property of $\mu_{\text{\normalfont eff}}(\cdot)$.

\section*{Acknowledgement}
BaoLuo Sun was supported by the National University of Singapore Start-Up Grant (R-155-000-203-133).  The authors thank Dr. Eric Tchetgen Tchetgen for helpful comments on a previous version of the manuscript, and Drs. Zhiqiang Tan and Heng Shu for help with the application data.

\bibliographystyle{Chicago}

\bibliography{refs}

\bigskip
\begin{center}
{\large\bf Appendix}
\end{center}

\subsection*{Proof of Theorem 1}

In the proof, we make use of the following equalities that for all square-integrable functions $m(X)$,
\begin{eqnarray*}
E \left\{\frac{(-1)^{1-Z}m(X)}{\lambda(Z|X)[\tau(1,X)-\tau(0,X)]}\biggr\vert R=1\right\}&=&0,\\
 E \left\{\frac{(-1)^{1-Z}m(X)\tau(Z,X)}{\lambda(Z|X)[\tau(1,X)-\tau(0,X)]}\biggr\vert R=1\right\}&=&E\{m(X)|R=1\}.
\end{eqnarray*}
Under assumptions 1--7 and suppressing the dependences of $\{h_0(\cdot),h_1(\cdot),g_0(\cdot),g_1(\cdot)\}$ on $(X,U)$,
\begin{eqnarray*}
&&E \left\{\frac{(-1)^{1-Z}}{\lambda(Z|X)}\frac{Y}{[\tau(1,X)-\tau(0,X)]}\biggr\vert R=1\right\}\\
&=& E \left\{\frac{(-1)^{1-Z}E(Y|D,Z,X,U,R=1)}{\lambda(Z|X)[\tau(1,X)-\tau(0,X)]}\biggr\vert R=1\right\}\\
&=&E \left\{\frac{(-1)^{1-Z}(h_0+h_1D)}{\lambda(Z|X)[\tau(1,X)-\tau(0,X)]}\biggr\vert R=1\right\}\\
&=&E\left\{\frac{(-1)^{1-Z}h_1(g_0+g_1Z)}{\lambda(Z|X)\{\tau(1,X)-\tau(0,X)\}}\biggr\vert R=1  \right\}+E\left\{\frac{(-1)^{1-Z}E(h_0|X,Z,R=1)}{\lambda(Z|X)\{\tau(1,X)-\tau(0,X)\}} \biggr\vert R=1 \right\}\\
&=& E\left\{\frac{(-1)^{1-Z}E[h_1|X,Z,R=1]\tau(X,Z)}{\lambda(Z|X)\{\tau(1,X)-\tau(0,X)\}} \biggr\vert R=1   \right\}+E\left\{\frac{(-1)^{1-Z}Z\text{cov}(g_1,h_1|X,Z,R=1)}{\lambda(Z|X)\{\tau(1,X)-\tau(0,X)\}}  \biggr\vert R=1  \right\}\\
&\phantom{=}&+E\left\{\frac{(-1)^{1-Z}\text{cov}(g_0, h_1|X,Z,R=1)}{\lambda(Z|X)\{\tau(1,X)-\tau(0,X)\}}  \biggr\vert R=1  \right\}+E\left\{\frac{(-1)^{1-Z}E(h_0|X,Z,R=1)}{\lambda(Z|X)\{\tau(1,X)-\tau(0,X)\}}   \biggr\vert R=1 \right\}\\
&=&  E\left\{\frac{(-1)^{1-Z}E[h_1|X,R=1]\tau(X,Z)}{\lambda(Z|X)\{\tau(1,X)-\tau(0,X)\}} \biggr\vert R=1   \right\}+E\left\{\frac{(-1)^{1-Z}Z\text{cov}(g_1,h_1|X,R=1)}{\lambda(Z|X)\{\tau(1,X)-\tau(0,X)\}}  \biggr\vert R=1  \right\}\\
&\phantom{=}&+E\left\{\frac{(-1)^{1-Z}\text{cov}(g_0, h_1|X,R=1)}{\lambda(Z|X)\{\tau(1,X)-\tau(0,X)\}}  \biggr\vert R=1  \right\}+E\left\{\frac{(-1)^{1-Z}E(h_0|X,R=1)}{\lambda(Z|X)\{\tau(1,X)-\tau(0,X)\}}   \biggr\vert R=1 \right\} \\
&=& E\left\{h_1|R=1 \right\}+E\left\{\frac{(-1)^{1-Z}Z\text{cov}(g_1,h_1|X,R=1)}{\lambda(Z|X)\{\tau(1,X)-\tau(0,X)\}}  \biggr\vert R=1  \right\},
\end{eqnarray*}
which equals the average treatment effect $\Delta$  if $\text{cov}(g_1,h_1|X,R=1)=0$ with probability 1. In addition, the propensity score $\tau(z,x)$ can be nonparametrically identified from $f(O)$ under assumptions 8 and 9. The proof is completed by noting that 
\begin{eqnarray*}
 E \left\{\frac{R}{q^{\dag}}\frac{(-1)^{1-Z}}{\lambda(Z|X)}\frac{Y}{[\tau(1,X)-\tau(0,X)]}\right\}= E \left\{\frac{(-1)^{1-Z}}{\lambda(Z|X)}\frac{Y}{[\tau(1,X)-\tau(0,X)]}\biggr \rvert R=1\right\}.
\end{eqnarray*}

\subsection*{Proof of Lemma 1}

\begin{eqnarray*}
&&E(Y|Z=z,X=x,R=1)\\
&=& \text{cov}(g_1,h_1|Z=z,X=x, R=1)z+\text{cov}(g_0,h_1 |Z=z,X=x,R=1)\\
&&\phantom{=}+E(h_1|Z=z,X=x,R=1)\tau(z,x)+E(h_0|Z=z,X=x,R=1)\\ 
&=& \text{cov}(g_1,h_1|X=x, R=1)z+\text{cov}(g_0,h_1 |X=x,R=1)\\
&&\phantom{=}+E(h_1|X=x,R=1)\tau(z,x)+E(h_0|X=x,R=1) \quad \text{by assumption 3}\\ 
&=&\mathcal{H}(x)\tau(z,x)+\omega(x)\quad \text{by assumption 6}.
\end{eqnarray*}

\subsection*{Proof of Lemma 2}
In the following, let $\bar{a}$ denote the probability limit of $\hat{a}$. Under standard theory for likelihood-based inference \citep{white1982maximum},
\begin{eqnarray*}
\hat{\alpha}-\bar{\alpha}&=&-\left\{\frac{\partial}{\partial \alpha}E[S_\pi(O;\alpha)]\bigr\vert_{\alpha=\bar{\alpha}}\right\}^{-1}\hat{E} \{S_\pi(O;\bar{\alpha})\}+o_p(n^{-1/2});\\
\hat{\psi}-\bar{\psi}&=&-\left\{\frac{\partial}{\partial \psi}E[S_\lambda(O;\psi)]\bigr\vert_{\psi=\bar{\psi}}\right\}^{-1}\hat{E} \{S_\lambda(O;\bar{\psi})\}+o_p(n^{-1/2});\\
\hat{\xi}-\bar{\xi}&=&-\left\{\frac{\partial}{\partial \xi}E[S_\tau(O;\xi)]\bigr\vert_{\xi=\bar{\xi}}\right\}^{-1}\hat{E} \{S_\tau(O;\bar{\xi})\}+o_p(n^{-1/2});\\
\hat{q}-\bar{q}&=&\hat{E} \{R-\bar{q}\}+o_p(n^{-1/2}),
\end{eqnarray*}
where $\{S_\pi, S_\lambda, S_\tau\}$ are the respective scores for the parametric models $\{\pi(z,x;\alpha),\lambda(z|x;\psi), \tau(z,x;\xi)\}$. Let $\delta_1=(\psi^T,\xi^T,q)^T$ denote the nuisance parameters in $\mathcal{M}_1$. By the asymptotic theory of M-estimators \citep{newey1994large, van2000asymptotic} and Taylor expansion, we obtain
\begin{eqnarray*}
\hat{\Delta}_1-\Delta&=& \hat{E}\{\mu_1(O;\Delta,\bar{\delta}_1)\}+(\hat{\psi}-\bar{\psi})^T\times \frac{\partial}{\partial \psi}\hat{E}\{\mu_1(O;\Delta,\delta_1)\}\bigr\vert_{\delta_1=\bar{\delta}_1}\\
&+&(\hat{\xi}-\bar{\xi})^T\times \frac{\partial}{\partial \xi}\hat{E}\{\mu_1(O;\Delta,\delta_1)\}\bigr\vert_{\delta_1=\bar{\delta}_1}+(\hat{q}-\bar{q})\times \frac{\partial}{\partial q}\hat{E}\{\mu_1(O;\Delta,\delta_1)\}\bigr\vert_{\delta_1=\bar{\delta}_1}+o_p(n^{-1/2}),
\end{eqnarray*}
so that $\hat{\Delta}_1-\Delta= \hat{E}\{\tilde{\mu}_1(O;\Delta,\bar{\delta}_1)\}+o_p(n^{1/2})$ where
\begin{eqnarray*}
\tilde{\mu}_1(O;\Delta,\bar{\delta}_1)&=&{\mu}_1(O;\Delta,\bar{\delta}_1)- \frac{\partial}{\partial \psi}E\{\mu_1(O;\Delta,\delta_1)\}\bigr\vert_{\delta_1=\bar{\delta}_1}\left\{\frac{\partial}{\partial \psi}E[S_\lambda(O;\psi)]\bigr\vert_{\psi=\bar{\psi}}\right\}^{-1}S_\lambda(O;\bar{\psi})\\
&-& \frac{\partial}{\partial \xi}E\{\mu_1(O;\Delta,\delta_1)\}\bigr\vert_{\delta_1=\bar{\delta}_1}\left\{\frac{\partial}{\partial \xi}E[S_\tau(O;\xi)]\bigr\vert_{\xi=\bar{\xi}}\right\}^{-1}S_\tau(O;\bar{\xi})\\
&+&\frac{\partial}{\partial q}E\{\mu_1(O;\Delta,\delta_1)\}\bigr\vert_{\delta_1=\bar{\delta}_1}\{R-\bar{q}\}+o_p(n^{-1/2}).
\end{eqnarray*}
Under $\mathcal{M}_1$, we have $\bar{\delta}_1={\delta}^{\dag}_1=(\psi^{\dag T},\xi^{\dag T},q^{\dag})^T$ and $E\{{\mu}_1(O;\Delta,{\delta}^{\dag}_1)\}=0$ by Theorem 1. It follows that $n^{1/2}(\hat{\Delta}_1-\Delta)\xrightarrow[]{d}N(0,\sigma^2_1)$ where $\sigma^2_1=E\{\tilde{\mu}^2_1(O;\Delta,{\delta}^{\dag}_1)\}$. 

Let $\delta_2=(\gamma^T,\eta^T,\xi^T,q)^T$ denote the nuisance parameters in $\mathcal{M}_2$. By Taylor expansion,
\begin{eqnarray*}
\hat{\Delta}_2-\Delta&=& \hat{E}\{{\mu}_2(O;\Delta,\bar{\gamma}_2, \bar{q})\}+(\hat{\gamma}_2-\bar{\gamma}_2)^T\times \frac{\partial}{\partial \gamma}\hat{E}\{\mu_2(O;\Delta,\gamma,q)\}\bigr\vert_{(\gamma,q)=(\bar{\gamma}_2,\bar{q})}\\
&+&(\hat{q}-\bar{q})\times \frac{\partial}{\partial q}\hat{E}\{\mu_2(O;\Delta,\gamma,q)\}\bigr\vert_{(\gamma,q)=(\bar{\gamma}_2,\bar{q})}+o_p(n^{-1/2}).
\end{eqnarray*}
Under $\mathcal{M}_2$, $(\bar{\xi},\bar{q})=(\xi^{\dag},q^{\dag})$, and at the true values $(\gamma^{\dag},\eta^{\dag})$, 
\begin{eqnarray*}
&& E\bigr\{\mathcal{G}_{v,w}(X,Z) \bigr\{R[Y-\mathcal{H}(X;\gamma^{\dag})\tau(Z,X; {\xi}^{\dag})-\omega(X;{\eta}^{\dag})]\\
&&\phantom{=}-(1-R)\mathcal{H}(X;\gamma^{\dag})[D-\tau(Z,X; {\xi}^{\dag})]\bigr\}\bigr\}\\
&=& E\bigr\{\mathcal{G}_{v,w}(X,Z)E \bigr\{R[Y-\mathcal{H}(X;\gamma^{\dag})\tau(Z,X; {\xi}^{\dag})-\omega(X;{\eta}^{\dag})]\notag\\
&&\phantom{=}-(1-R)\mathcal{H}(X;\gamma^{\dag})[D-\tau(Z,X; {\xi}^{\dag})]\mid Z,X\bigr\}\bigr\}\\
&=&E\bigr\{\mathcal{G}_{v,w}(X,Z) \bigr\{[E(Y|Z,X,R=1)-\mathcal{H}(X;\gamma^{\dag})\tau(Z,X; {\xi}^{\dag})-\omega(X;\eta^{\dag})]\pi(Z,X)\notag\\
&&\phantom{=}-\mathcal{H}(X;\gamma^{\dag})[E(D|Z,X,R=0)-\tau(Z,X; {\xi}^{\dag})](1-\pi(Z,X))\bigr\}\bigr\}\\
&=&0,
\end{eqnarray*}
so that under standard regularity conditions for M-estimation {\citep{newey1994large, van2000asymptotic}} $\bar{\delta}_2=\delta_2^{\dag}$. We have $E\{{\mu}_2(O;\Delta,{\gamma}^{\dag}, {q}^{\dag})\}=0$ by definition.  The asymptotic distribution of $n^{1/2}(\hat{\Delta}_2-\Delta)$ follows from the previous Taylor expansions by Slutsky's Theorem and the Central Limit Theorem. The  expansion for $\hat{\Delta}_3-\Delta$ can be proven similarly; we note that under $\mathcal{M}_3$, $(\bar{\alpha},\bar{q})=(\alpha^{\dag},q^{\dag})$, and at the true values $(\gamma^{\dag},\eta^{\dag})$, 
\begin{eqnarray*}
&&E\left\{ \mathcal{G}_{v,w}(X,Z) \biggr\{R[Y-\omega(X;{\eta}^{\dag})]-\frac{(1-R)\pi(Z,X;{\alpha}^{\dag})}{1-\pi(Z,X;{\alpha}^{\dag})}\mathcal{H}(X;\gamma^{\dag})D  \biggr\} \right\}\\
&=&E\left\{ \mathcal{G}_{v,w}(X,Z) E\biggr\{R[Y-\omega(X;{\eta}^{\dag})]-\frac{(1-R)\pi(Z,X;{\alpha}^{\dag})}{1-\pi(Z,X;{\alpha}^{\dag})}\mathcal{H}(X;\gamma^{\dag})D  \biggr \vert Z,X\biggr\} \right\}\\
&=& E  \biggr\{\mathcal{G}_{v,w}(X,Z) \biggr\{\pi(Z,X)[E(Y|Z,X,R=1)-\omega(X;\eta^{\dag})]\\
&&\phantom{=}-\pi(Z,X;{\alpha}^{\dag})\mathcal{H}(X;\gamma^{\dag})E(D|Z,X,R=0)  \biggr\} \biggr\}\\
&=& E  \biggr\{\mathcal{G}_{v,w}(X,Z) \biggr\{\pi(Z,X)\mathcal{H}(X)\tau(Z,X)\\
&&\phantom{=}-\pi(Z,X;{\alpha}^{\dag})\mathcal{H}(X;\gamma^{\dag})E(D|Z,X,R=0)  \biggr\} \biggr\}\\
&=&0.
\end{eqnarray*}

\subsection*{Proof of Theorem 2}

We closely follow the structure of semiparametric efficiency bound derivation of \citet{newey1990semiparametric}, \citet{bickel1993efficient} and \citet{chen2008semiparametric}. Consider a parametric path $t$  for the density of the observed data,  $f_t(O)=q^R_t(1-q_t)^{1-R}f_t(V|R=1)^Rf_t(V|R=0)^{1-R}f_t(Y|V,R=1)^Rf_t(D|V)^{1-R},$ where $q_t=\text{pr}_t(R=1)$. We aim to derive the unique influence function $\mu_{\text{\normalfont eff}}(O)$ under $\mathcal{M}_{\text{np}}$ such that $E\{\mu_{\text{\normalfont eff}}(O)\}=0$ and pathwise differentiability holds: $$\partial \Delta_t /\partial t=E_{t}\left\{\mu_{\text{\normalfont eff}}(O)S_{t}(O)\right\},$$ where 
\begin{eqnarray*}
S_{t}(O)&=&\partial  \log f_t(O)/\partial t\\
&=&\alpha(R-q_t)+(1-R)S_t(V|R=0)+R S_t(V|R=1)+RS_t(Y|V,R=1)+(1-R) S_t(D|V). 
\end{eqnarray*}
 
Following the proof for Theorem 1, $$\Delta_t= E_t \left\{\frac{R}{q_t}\frac{(-1)^{1-Z}Y}{\lambda_t(Z|X)[\tau_t(1,X)-\tau_t(0,X)]}\right\}=E_t \left\{\frac{(-1)^{1-Z}Y}{\lambda_t(Z|X)[\tau_t(1,X)-\tau_t(0,X)]}\biggr \vert R=1\right\}.$$ Differentiate the integral on the right hand side with respect to $t$ yields
\begin{eqnarray*}
\frac{\partial \Delta_t}{\partial t}&=&E_t \left\{\frac{(-1)^{1-Z}YS_t(Y,D,V|R=1)}{\lambda_t(Z|X)[\tau_t(1,X)-\tau_t(0,X)]}\biggr \vert R=1\right\}-E_t \left\{\frac{(-1)^{1-Z}YS_t(Z|X,R=1)}{\lambda_t(Z|X)[\tau_t(1,X)-\tau_t(0,X)]}\biggr \vert R=1\right\}\\
\phantom{=}&-& E_t \left\{\frac{(-1)^{1-Z}Y\frac{\partial [\tau_t(1,X)-\tau_t(0,X)]}{\partial t}}{\lambda_t(Z|X)[\tau_t(1,X)-\tau_t(0,X)]^2}\biggr \vert R=1\right\}\\
\phantom{=}&\equiv& A_1+A_2+A_3.
\end{eqnarray*}
 Consider the terms separately:
\begin{eqnarray*}
A_2&=&-E_t \left\{\frac{(-1)^{1-Z}YS_t(Z|X,R=1)}{\lambda_t(Z|X)[\tau_t(1,X)-\tau_t(0,X)]}\biggr \vert R=1\right\}\\
&=&-E_t\left\{ \left\{ 
\begin{array}{c}
E_t\left[\frac{(-1)^{1-Z}Y}{\lambda_t(Z|X)[\tau_t(1,X)-\tau_t(0,X)]} \biggr \rvert V, R=1\right]  \\ 
-E\left[ \frac{(-1)^{1-Z}Y}{\lambda_t(Z|X)[\tau_t(1,X)-\tau_t(0,X)]}  \biggr \rvert X,R=1\right] 
\end{array}%
\right\} S_{t}(Z|X,R=1)\Biggr \vert R=1\right\}  \\
&=&-E_t\left\{ \left\{ 
\begin{array}{c}
E_t\left[\frac{(-1)^{1-Z}Y}{\lambda_t(Z|X)[\tau_t(1,X)-\tau_t(0,X)]}  \biggr \rvert V, R=1\right]  \\ 
-E_t\left[ \frac{(-1)^{1-Z}Y}{\lambda_t(Z|X)[\tau_t(1,X)-\tau_t(0,X)]}  \biggr \rvert X, R=1\right] 
\end{array}%
\right\} S_{t}(Y,D,V|R=1)\Biggr \vert R=1\right\}  \\
&=&-E_t\biggr\{ \left\{\frac{(-1)^{1-Z}\mathcal{H}(X)\tau_t(Z,X)}{\lambda_t(Z|X)\{\tau_t(1,X)-\tau_t(0,X)\}}
+\frac{(-1)^{1-Z}\omega(X)}{\lambda_t(Z|X)\{\tau_t(1,X)-\tau_t(0,X)\}} -\mathcal{H}(X)\right\}\times\\
&& S_{t}(Y,D,V|R=1)\biggr \vert R=1\biggr\}\\
&=&-E_t\biggr\{ \left\{\frac{(-1)^{1-Z}\mathcal{H}(X)\tau_t(Z,X)}{\lambda_t(Z|X)\{\tau_t(1,X)-\tau_t(0,X)\}}
+\frac{(-1)^{1-Z}\omega(X)}{\lambda_t(Z|X)\{\tau_t(1,X)-\tau_t(0,X)\}} -\mathcal{H}(X)+\Delta\right\}\times\\
&& S_{t}(Y,D,V|R=1)\biggr \vert R=1\biggr\},
\end{eqnarray*}

\begin{eqnarray*}
A_3&=&- E_t \left\{\frac{(-1)^{1-Z}Y\frac{\partial [\tau_t(1,X)-\tau_t(0,X)]}{\partial t}}{\lambda_t(Z|X)[\tau_t(1,X)-\tau_t(0,X)]^2}\biggr \vert R=1\right\}\\
&=&-  E_t \left\{\frac{(-1)^{1-Z}YE(DS_t(D|Z=1,X,R=1)|Z=1,X,R=1)}{\lambda_t(Z|X)[\tau_t(1,X)-\tau_t(0,X)]^2}\biggr \vert R=1\right\} \\
&&+E_t \left\{\frac{(-1)^{1-Z}YE(DS_t(D|Z=0,X,R=1)|Z=0,X,R=1)}{\lambda_t(Z|X)[\tau_t(1,X)-\tau_t(0,X)]^2}\biggr \vert R=1\right\} \\
&=&- \left\{ E_t \left\{\frac{\mathcal{H}(X)ZDS_t(D|V,R=1)}{\lambda_t(Z|X)[\tau_t(1,X)-\tau_t(0,X)]}\biggr \vert R=1\right\} -E_t \left\{\frac{\mathcal{H}(X)(1-Z)DS_t(D|V,R=1)}{\lambda_t(Z|X)[\tau_t(1,X)-\tau_t(0,X)]}\biggr \vert R=1\right\} \right\}\\
&=&-  E_t \left\{\frac{\mathcal{H}(X)(-1)^{1-Z}DS_t(D|V,R=1)}{\lambda_t(Z|X)[\tau_t(1,X)-\tau_t(0,X)]}\biggr \vert R=1\right\}\\
&=& - E_t \left\{\frac{\mathcal{H}(X)(-1)^{1-Z}[D-\tau_t(Z,X)]S_t(D|V,R=1)}{\lambda_t(Z|X)[\tau_t(1,X)-\tau_t(0,X)]}\biggr \vert R=1\right\}\\
&=& - E_t \left\{\frac{\mathcal{H}(X)(-1)^{1-Z}[D-\tau_t(Z,X)]S_t(Y,D,V|R=1)}{\lambda_t(Z|X)[\tau_t(1,X)-\tau_t(0,X)]}\biggr \vert R=1\right\}.
\end{eqnarray*}
Combining the terms $A_1$--$A_3$, 
\begin{eqnarray*}
\frac{\partial \Delta_t}{\partial t}&=&  E_t\left\{\left[\frac{\left( -1\right)^{1-Z}\left\{ Y-\omega(X)-\mathcal{H}(X)D \right\}}{\lambda_t(Z|X)\left[ \tau_t(1,X)-\tau_t(0,X)\right] }+\mathcal{H}(X)-\Delta\right]S_t(Y,D,V|R=1)\biggr \vert R=1\right\}\\
&\equiv& E_t\{\varphi(Y,D,V)S_t(Y,D,V|R=1)|R=1\}.
\end{eqnarray*}

Let $\varphi(Y,D,V)=\varphi_1(Y,V)-\varphi_2(D,V)$ where 
\begin{eqnarray*}
\varphi_1(Y,V)&=&\frac{\left( -1\right)^{1-Z}Y}{\lambda_t(Z|X)\left[ \tau_t(1,X)-\tau_t(0,X)\right] };\\
\varphi_2(D,V)&=&\frac{\left( -1\right)^{1-Z}\{\omega(X)+\mathcal{H}(X)D\}}{\lambda_t(Z|X)\left[ \tau_t(1,X)-\tau_t(0,X)\right] }-\mathcal{H}(X)+\Delta.
\end{eqnarray*}
Then
\begin{eqnarray*}
\frac{\partial \Delta_t}{\partial t}&= &E_t\{\varphi(Y,D,V)S_t(Y,D,V|R=1)|R=1\}\\
&= & E_t\left[\varphi_1(Y,V)S_t(Y,V|R=1)|R=1 \right]-E_t\left[\varphi_2(D,V)S_t(D,V|R=1)|R=1\right]  \\
&\equiv& B_1-B_2
\end{eqnarray*}
We note that 
\begin{eqnarray*}
B_1&=& E_t\left[\varphi_1(Y,V)S_t(Y|V,R=1) |R=1\right]+E_t\left[\varphi_1(Y,V)S_t(V|R=1) |R=1\right]\\
&=& E_t \left\{\frac{R}{q_t}[\varphi_1(Y,V)-E(\varphi_1|V,R=1)]S_t(Y|V,R=1) \right\}\\
&\phantom{=}&+ E_t\left\{\frac{E_t(\varphi_1|V,R=1)}{q_t}RS_t(V|R=1)  \right\}\\
&=&E_t \left\{\frac{R}{q_t}[\varphi_1(Y,V)-E(\varphi_1|V,R=1)]S_t(O) \right\}+ E_t\left\{\frac{E_t(\varphi_1|V,R=1)}{q_t}RS_t(O)  \right\},\\
\text{and similarly}\\
B_2&=& E_t\left[\varphi_2(D,V)S_t(D|V) |R=1\right]+E_t\left[\varphi_2(D,V)S_t(V|R=1) |R=1\right]\\
&=& E_t \left\{\frac{1-R}{q_t}\frac{\pi_t(V)}{1-\pi_t(V)}[\varphi_2(D,V)-E(\varphi_2|V)]S_t(D|V) \right\}+ E_t\left\{\frac{E_t(\varphi_2|V)}{q_t}RS_t(V|R=1)  \right\}\\
&=&E_t \left\{\frac{1-R}{q_t}\frac{\pi_t(V)}{1-\pi_t(V)}[\varphi_2(D,V)-E(\varphi_2|V)]S_t(O) \right\}+ E_t\left\{\frac{E_t(\varphi_2|V)}{q_t}RS_t(O)  \right\},
\end{eqnarray*}
where $q_t=\int \pi_t(v)f_t(v)dv$. Therefore $\frac{\partial \Delta_t}{ \partial t}=E_t\left\{\mu_{\text{\normalfont eff}}(O)S_t(O) \right\},$ where
\begin{eqnarray*}
\mu_{\text{\normalfont eff}}(O;\Delta)&=&\frac{R}{q_t}[\varphi_1(Y,V)-E(\varphi_1|V,R=1)]-\frac{1-R}{q_t}\frac{\pi_t(V)}{1-\pi_t(V)}[\varphi_2(D,V)-E(\varphi_2|V)]\\
&&+\frac{R}{q_t}[E(\varphi_1|V,R=1)-E(\varphi_2|V)].
\end{eqnarray*}
It it straightforward to verify that $E_t\{\mu_{\text{\normalfont eff}}(O;\Delta)\}=0.$ It follows by standard semiparametric efficiency theory that $\mu_{\text{\normalfont eff}}(O;\Delta)$ is the unique (and hence also efficient) influence function, and the semiparametric efficiency bound for all regular and asymptotically linear estimators of $\Delta$ in $\mathcal{M}_{\text{np}}$ is $E\{\mu^2_{\text{\normalfont eff}}(O;\Delta)\}$.

\subsection*{Proof of Lemma 3}
Let $\delta=(\eta^T,\gamma^T,\psi^T,\xi^T,\alpha^T,q)^T$ denote the nuisance parameters. By the asymptotic theory of M-estimators \citep{van2000asymptotic} and Taylor expansion, we obtain
\begin{eqnarray*}
\hat{\Delta}_{\text{mul}}-\Delta&=& \hat{E}\{\mu_{\text{\normalfont eff}}(O;\Delta,\bar{\delta})\}+(\hat{\delta}-\bar{\delta})^T\times \frac{\partial}{\partial \delta}\hat{E}\{\mu_{\text{\normalfont eff}}(O;\Delta,\delta)\}\bigr\vert_{\delta=\bar{\delta}}+o_p(n^{-1/2}).
\end{eqnarray*}
It suffices to show that ${E}\{\mu_{\text{\normalfont eff}}(O;\Delta,\bar{\delta})\}=0$ in the union model $\cup_{j=1}^3 \mathcal{M}_j$.

Under $\mathcal{M}_1$, we have $(\bar{\psi},\bar{\xi},\bar{q})=(\psi^{\dag },\xi^{\dag },q^{\dag})$ and
\begin{eqnarray*}
&&E\{\mu_{\text{\normalfont eff}}(O;\Delta, \bar{\eta},\bar{\gamma},\psi^{\dag},\xi^{\dag},\bar{\alpha},q^{\dag})\}\\
&=& E \Biggr \{\frac{\left( -1\right)^{1-Z} }{\lambda(Z|X;\psi^{\dag})}\times \frac{\frac{R}{{q}^{\dag}}Y }{[\tau(1,X;\xi^{\dag})-\tau(0,X; \xi^{\dag})]} \Biggr \}\\
&\phantom{=}&-E \Biggr \{\frac{\left( -1\right)^{1-Z} }{\lambda(Z|X;\psi^{\dag})}\times \frac{\frac{R}{{q}^{\dag}}[\mathcal{H}(X;\bar{\gamma}) 
\tau(Z,X;{\xi}^{\dag})+\omega(X;\bar{\eta})] }{[\tau(1,X;\xi^{\dag})-\tau(0,X; \xi^{\dag})]} \Biggr \}\notag\\
&\phantom{=}&- E \Biggr \{\frac{\left( -1\right)^{1-Z} }{\lambda(Z|X;\psi^{\dag})}\times \frac{\frac{1-R}{{q}^{\dag}}\frac{\pi(Z,X;\bar{\alpha})}{1-\pi(Z,X;\bar{\alpha})}\mathcal{H}(X;\bar{\gamma}) [D-\tau(Z,X;{\xi}^{\dag})] }{[\tau(1,X; {\xi}^{\dag})-\tau(0,X; {\xi}^{\dag})]}\Biggr \}+E\biggr \{\frac{R}{{q}^{\dag}}[\mathcal{H}(X;\bar{\gamma})]\biggr \}\\
&=& E \Biggr \{\frac{\left( -1\right)^{1-Z} }{\lambda(Z|X;\psi^{\dag})}\times \frac{\frac{R}{{q}^{\dag}}Y }{[\tau(1,X;\xi^{\dag})-\tau(0,X; \xi^{\dag})]} \Biggr \}-E \Biggr \{\frac{R}{{q}^{\dag}}\mathcal{H}(X;\bar{\gamma}) \Biggr \}\notag\\
&\phantom{=}&- E \Biggr \{\frac{\left( -1\right)^{1-Z} }{\lambda(Z|X;\psi^{\dag})}\times \frac{\frac{1-\pi(Z,X)}{{q}^{\dag}}\frac{\pi(Z,X;\bar{\alpha})}{1-\pi(Z,X;\bar{\alpha})}\mathcal{H}(X;\bar{\gamma}) [E(D|Z,X,R=0)-\tau(Z,X;{\xi}^{\dag})] }{[\tau(1,X; {\xi}^{\dag})-\tau(0,X; {\xi}^{\dag})]}\Biggr \}\notag\\
&\phantom{=}&+E\biggr \{\frac{R}{{q}^{\dag}}[\mathcal{H}(X;\bar{\gamma})-\Delta]\biggr \}\\
&=& E \Biggr \{\frac{\left( -1\right)^{1-Z} }{\lambda(Z|X;\psi^{\dag})}\times \frac{\frac{R}{{q}^{\dag}}Y }{[\tau(1,X;\xi^{\dag})-\tau(0,X; \xi^{\dag})]} \Biggr \}-\Delta\\
&=&0,
\end{eqnarray*}
by Theorem 1.

Under $\mathcal{M}_2$, $(\bar{\xi},\bar{q})=(\xi^{\dag },q^{\dag})$ and at the true values $(\gamma^{\dag},\eta^{\dag})$,
\begin{eqnarray*}
&&{E} \Biggr \{ \mathcal{G}_{v,w}(X,Z) \biggr\{R[Y-\mathcal{H}(X;\gamma^{\dag })\tau(Z,X; {\xi}^{\dag })-\omega(X; {\eta}^{\dag })]\notag\\
&\phantom{=}&-\frac{(1-R)\pi(Z,X;\bar{\alpha})}{1-\pi(Z,X;\bar{\alpha})}\mathcal{H}(X;\gamma^{\dag })[D-\tau(Z,X; {\xi}^{\dag })]  \biggr\} \biggr\}\\
&=&{E} \Biggr \{ \mathcal{G}_{v,w}(X,Z) E\biggr\{R[Y-\mathcal{H}(X;\gamma^{\dag })\tau(Z,X; {\xi}^{\dag })-\omega(X; {\eta}^{\dag })]\notag\\
&\phantom{=}&-\frac{(1-R)\pi(Z,X;\bar{\alpha})}{1-\pi(Z,X;\bar{\alpha})}\mathcal{H}(X;\gamma^{\dag })[D-\tau(Z,X; {\xi}^{\dag })]  \biggr\vert Z,X\biggr\} \biggr\}\\
&=&  {E} \Biggr \{ \mathcal{G}_{v,w}(X,Z) \biggr\{\pi(Z,X)[E(Y|Z,X,R=1)-\mathcal{H}(X;\gamma^{\dag})\tau(Z,X; {\xi}^{\dag})-\omega(X; \eta^{\dag})]\notag\\
&-&\frac{(1-\pi(Z,X))\pi(Z,X;\bar{\alpha})}{1-\pi(Z,X;\bar{\alpha})}\mathcal{H}(X;\gamma^{\dag})[E(D|Z,X,R=1)-\tau(Z,X; {\xi}^{\dag})]  \biggr\} \biggr\}=0.
\end{eqnarray*}
In addition,
\begin{eqnarray*}
&&E\{\mu_{\text{\normalfont eff}}(O;\Delta, {\eta}^{\dag},{\gamma}^{\dag},\bar{\psi},\xi^{\dag},\bar{\alpha},q^{\dag})\}=E\{E\{\mu_{\text{\normalfont eff}}(O;\Delta, {\eta}^{\dag},{\gamma}^{\dag},\bar{\psi},\xi^{\dag},\bar{\alpha},q^{\dag})|Z,X\}\}\\
&=&  E \biggr \{\frac{\left( -1\right)^{1-Z} }{\lambda(Z|X;,\bar{\psi})}\times\frac{\frac{\pi(Z,X)}{{q}^{\dag}}[ E(Y|Z,X,R=1)-\mathcal{H}(X;{\gamma}^{\dag}) 
\tau(Z,X;\xi^{\dag})-\omega(X;\eta^{\dag})] }{[\tau(1,X; \xi^{\dag})-\tau(0,X; \xi^{\dag})]} \biggr \}\\
&\phantom{=}&- E \biggr \{\frac{\left( -1\right)^{1-Z} }{\lambda(Z|X;,\bar{\psi})}\times\frac{\frac{1-\pi(Z,X)}{{q}^{\dag}}\frac{\pi(Z,X;\bar{\alpha})}{1-\pi(Z,X;\bar{\alpha})}\mathcal{H}(X;{\gamma}^{\dag}) [E(D|Z,X,R=0)-\tau(Z,X;\xi^{\dag})] }{[\tau(1,X; \xi^{\dag})-\tau(0,X; \xi^{\dag})]} \biggr \} \\
&\phantom{=}&+E\biggr \{\frac{R}{{q}^{\dag}}[\mathcal{H}(X;{\gamma}^{\dag})-\Delta]\biggr \}\\
&=&E\{\mathcal{H}(X;{\gamma}^{\dag})|R=1\}-\Delta = 0.
\end{eqnarray*}

Under $\mathcal{M}_3$, $(\bar{\alpha},\bar{q})=(\alpha^{\dag },q^{\dag})$ and at the true values $(\gamma^{\dag},\eta^{\dag})$,

\begin{eqnarray*}
&&{E} \Biggr \{ \mathcal{G}_{v,w}(X,Z) \biggr\{R[Y-\mathcal{H}(X;\gamma^{\dag })\tau(Z,X; \bar{\xi})-\omega(X; {\eta}^{\dag })]\notag\\
&\phantom{=}&-\frac{(1-R)\pi(Z,X;{\alpha}^{\dag })}{1-\pi(Z,X;{\alpha}^{\dag })}\mathcal{H}(X;\gamma^{\dag })[D-\tau(Z,X;\bar{\xi})]  \biggr\} \biggr\}\\
&=&{E} \Biggr \{ \mathcal{G}_{v,w}(X,Z) E\biggr\{R[Y-\mathcal{H}(X;\gamma^{\dag })\tau(Z,X; \bar{\xi})-\omega(X; {\eta}^{\dag })]\notag\\
&\phantom{=}&-\frac{(1-R)\pi(Z,X;{\alpha}^{\dag })}{1-\pi(Z,X;{\alpha}^{\dag })}\mathcal{H}(X;\gamma^{\dag })[D-\tau(Z,X; \bar{\xi})]  \biggr\vert Z,X\biggr\} \biggr\}\\
&=&  \hat{E} \Biggr \{\mathcal{G}_{v,w}(X,Z)   \biggr\{\pi(Z,X)\mathcal{H}(X;\gamma^{\dag})[\tau(Z,X)-\tau(Z,X;\bar{\xi})]  \notag\\
&-&\pi(Z,X;\alpha^{\dag})\mathcal{H}(X;\gamma^{\dag})[\tau(Z,X)-\tau(Z,X; \bar{\xi})]  \biggr\} \biggr\}=0.
\end{eqnarray*}
In addition,
\begin{eqnarray*}
&&E\{\mu_{\text{\normalfont eff}}(O;\Delta,, {\eta}^{\dag},{\gamma}^{\dag},\bar{\psi},\bar{\xi},{\alpha}^{\dag},q^{\dag})\}=E\{E\{\mu_{\text{\normalfont eff}}(O;\Delta,{\eta}^{\dag},{\gamma}^{\dag},\bar{\psi},\bar{\xi},{\alpha}^{\dag},q^{\dag})|Z,X\}\}\\
&=&  E \biggr \{\frac{\left( -1\right)^{1-Z} }{\lambda(Z|X;,\bar{\psi})}\times\frac{\frac{\pi(Z,X)}{{q}^{\dag}}[ E(Y|Z,X,R=1)-\mathcal{H}(X;{\gamma}^{\dag}) 
\tau(Z,X; \bar{\xi})-\omega(X;\eta^{\dag})] }{[\tau(1,X; \bar{\xi})-\tau(0,X;  \bar{\xi})]} \biggr \}\\
&\phantom{=}&- E \biggr \{\frac{\left( -1\right)^{1-Z} }{\lambda(Z|X;,\bar{\psi})}\times\frac{\frac{\pi(Z,X;\alpha^{\dag})}{{q}^{\dag}}\mathcal{H}(X;{\gamma}^{\dag}) [E(D|Z,X,R=0)-\tau(Z,X; \bar{\xi})] }{[\tau(1,X;  \bar{\xi})-\tau(0,X;  \bar{\xi})]} \biggr \} \\
&\phantom{=}&+E\biggr \{\frac{R}{{q}^{\dag}}[\mathcal{H}(X;{\gamma}^{\dag})-\Delta]\biggr \}\\
&=&  E \biggr \{\frac{\left( -1\right)^{1-Z} }{\lambda(Z|X;,\bar{\psi})}\times\frac{\frac{\pi(Z,X)}{{q}^{\dag}}\mathcal{H}(X;{\gamma}^{\dag}) [\tau(Z,X) -\tau(Z,X; \bar{\xi})] }{[\tau(1,X; \bar{\xi})-\tau(0,X;  \bar{\xi})]} \biggr \}\\
&\phantom{=}&- E \biggr \{\frac{\left( -1\right)^{1-Z} }{\lambda(Z|X;,\bar{\psi})}\times\frac{\frac{\pi(Z,X;\alpha^{\dag})}{{q}^{\dag}}\mathcal{H}(X;{\gamma}^{\dag}) [\tau(Z,X)-\tau(Z,X; \bar{\xi})] }{[\tau(1,X;  \bar{\xi})-\tau(0,X;  \bar{\xi})]} \biggr \} \\
&\phantom{=}&+E\biggr \{\frac{R}{{q}^{\dag}}[\mathcal{H}(X;{\gamma}^{\dag})-\Delta]\biggr \}\\
&=&E\{\mathcal{H}(X;{\gamma}^{\dag})|R=1\}-\Delta = 0.
\end{eqnarray*}

The last claim in Lemma 4 follows by noting that under the intersection submodel $\left\{\cap_{j=1}^3\mathcal{M}_{j}\right\}$, $\bar{\delta}=\delta^{\dag}$ and $\frac{\partial}{\partial \delta}\hat{E}\{\mu_{\text{\normalfont eff}}(O;\Delta,\delta)\}\bigr\vert_{\delta={\delta}^{\dag}}=o_p(1)$ so that
\begin{eqnarray*}
\hat{\Delta}_{\text{mul}}-\Delta&=& \hat{E}\{\mu_{\text{\normalfont eff}}(O;\Delta,{\delta}^{\dag})\}+o_p(n^{-1/2}).
\end{eqnarray*}

\end{document}